\documentclass[twocolumn,prl,amsmath,amssymb,showpacs,
superscriptaddress,floatfix]{revtex4}

\usepackage{graphicx}
\usepackage {amsmath}

\DeclareMathOperator{\sign}{sign}
\begin{document}

\title{Andreev quantum dots for spin manipulation}

\author{Nikolai~M. Chtchelkatchev}
\affiliation{L.D.\ Landau Institute for Theoretical Physics,
Russian Academy of Sciences, 117940 Moscow, Russia}

\affiliation{Institute for High Pressure Physics, Russian Academy
of Sciences, Troitsk 142092, Moscow Region, Russia}

\author{Yu. V. Nazarov}
\affiliation{Department of
Nanoscience and DIMES, Delft University of Technology, Lorentzweg
1, 2628 CJ Delft, the Netherlands}

\date{\today}

\begin{abstract}
We investigate the feasibility of manipulating individual
spin in a superconducting junction where
Bogolyubov quasiparticles
can be trapped in discrete Andreev
levels.  We call this system Andreev Quantum Dot (AQD)
to be contrasted with a common semiconductor quantum dot.
We show that AQD can be brought into a spin-1/2
state. The coupling between spin and superconducting current
facilitate manipulation and measurement of this state.
We demonstrate that one can
operate two inductively  coupled AQD's as a XOR gate,
this enables quantum computing applications.
\end{abstract}


\maketitle

Manipulation and operation of individual quantum systems and
arrays of such systems, so-called ``quantum machines'' is now in
focus of both experimental and theoretical research
\cite{1Vion-Devoret,2Yu-Jozsa}. The progress in quantum computing
algorithms \cite{3Ekert-Jozsa} has demonstrated potential
applicability of quantum mechanics thus stimulating various
proposals to implement arrays of operational two-state systems
(qubits) in solid-state
\cite{4Loss-DiVincenzo,5Recher-Levy,6Shnirman-Hermon,7Averin,
8Mooij-Lloyd,9Kane,10Makhlin-Schnirman,11Ioffe-Blatter,
12Nakamura-Tsai,13Ioffe}. Many proposals concern quantum dots. The
quantum dots are often referred to as artificial atoms since they
confine a discrete number of particles that occupy discrete
quantum states. In contrast to atoms, the properties of quantum
dots can be tuned and their charge and spin degrees of freedom can
be controlled. This would allow for quantum manipulation. An
interesting and elaborated proposal
\cite{4Loss-DiVincenzo,5Recher-Levy} utilizes spin states of
semiconductor quantum dots. However, the complexity of the
manipulation schemes proposed and severe difficulties with the
read-out of these spin states \cite{5Recher-Levy,14Recher} drives
one to think of alternatives.

Below we present an alternative scheme for individual spin
manipulation. We concentrate on sufficiently resistive
superconducting constrictions where individual Bogolybov
quasiparticles can be trapped in {\it discrete} Andreev bound
states. We refer to such system as Andreev Quantum Dots (AQD). An
AQD resembles a common quantum dot as long as discreetness of a
(quaisi)particle number, spectrum and spin is concerned. Albeit in
contrast to a common quantum dot the charge of the AQD is not
fixed. This allows for superconducting current in the constriction
and makes electron-electron interaction negligible.

We propose to utilize
spin states of the AQD's. We show that an AQD can be brought to
the state with spin-1/2 that persist over a long time. It is
important that the spin direction in this state determines the
superconducting current in the constriction, thus solving the
read-out problem. We demonstrate that the spin state of a single
AQD can be manipulated. Further, the two dots can be inductively
coupled to make a XOR quantum gate. Quantum information theory
\cite{3Ekert-Jozsa}
proves  that this enables one to build a
universal quantum computer.

The AQD can be formed in any constriction between two
superconducting leads that have a gap in energy spectrum. If an
electron with the energy below the gap tries to escape to the bulk
of a superconductor, it is reflected back as a hole (Andreev
reflection \cite{Andreev}), which also can not escape due to the
same reason. So that, the junction \textit{confines}
quasiparticles that are coherent mixtures of electron and holes.
Their discrete energy levels and eigenfunctions are determined by
Bogoliubov equations (BdG) \cite{dG}. It is sometimes forgotten
that these equations do possess a spin structure. Bogoliubov
eigenfunctions are made of two spinors \cite{Landau3}
$u^{\alpha},v^{\alpha}$ [coefficients of the Bogoliubov
transformation \cite{dG}
$\Psi(r,\sigma)=\sum_n(u_n(r,\sigma)\gamma_n+
g^{\sigma\mu}v_n^*(r,\mu)\gamma_n^{\dag})$] that satisfy
\begin{gather}
\label{BdG}
\begin{aligned}
\varepsilon u^{\alpha}&={\hat H}^{\alpha}_{\;\;\beta} u^{\beta}+
\hat \Delta v^{\alpha}\\
\varepsilon v^{\alpha}&=-[\hat
H^{*}]^{\alpha}_{\;\;\beta}v^{\beta}+{\hat\Delta}^* u^{\alpha},\;
\end{aligned}
\end{gather}
Here ``hat'' denotes an operator over orbital degrees of freedom.
We make explicit the spin structure of the single-particle
Hamiltonian $H$ and pair potential $\Delta$, $g_{\alpha\beta}
\equiv i\sigma^y$ being metric tensor in spinor space
\cite{Landau3}, $(\hat H^*)^{\alpha}_{\;\;\beta}\equiv
g^{\nu\alpha}(\hat H^{\nu}_{\;\;\mu})^*g_{\mu\beta}$. By virtue of
Eq. \eqref{BdG} quasiparticle energy levels always come in pairs:
each eigenstate with energy $\varepsilon$ has a counterpart with
energy $-\varepsilon$. This is due to a double-counting: there are
two quasiparticle eigenfunctions per each state of $H$. Should
$\hat H$ possess no spin structure, Andreev levels are
\textit{spin-degenerate}. For many problems that do {\it not}
involve spin, one can avoid the double counting by considering one
\textit{non-degenerate} level. This technical trick does not
correspond to original formulation of superconductivity theory
\cite{17_1/2,dG}, neither it gives the correct description of spin
in superconductors.

We concentrate on a short constriction, such that the typical time
for an electron $\tau_{_{\rm flight}}$ to traverse the junction
satisfies the condition $\tau_{_{\rm flight}} \ll \hbar/\Delta$.
In the limit $\tau_{_{\rm flight}}\Delta/\hbar \rightarrow 0$, and
in the absence of magnetic field Andreev levels are
\textit{spin-degenerate} and can be universally expressed
\cite{16Beenakker} through eigenvalues $T_n$ of the transmission
matrix square, $\varepsilon_{n_1,n_2;\,\sigma}=\sign(n_2)\Delta
\sqrt{1-T_{n_1}\sin^2(\varphi/2)}$. Here the integer index $n_1$
labels orbital channels, $n_2=\pm 1$, $\sigma=\pm 1$ is spin-index
and $\varphi$ stands for the superconducting phase difference
between the leads.

Andreev levels that are relevant for electron transport, and for
manipulation of spin states, originate from $T_n \lesssim 1$.
These levels are distributed in energy strip
$\Delta|\cos(\varphi/2)|< |\varepsilon|< \Delta$. Their typical
spacing is given by $\delta E\sim\Delta G_Q R$,  $R$ being the
normal state resistance of the constriction,$G_Q$ being the
conductance quantum \cite{15Imry}. This estimation is valid for
any sufficiently disordered constriction except tunnel junctions
for which all $T_n\ll 1$. To give an estimation, for $R \approx$
10 Ohm and $\Delta\approx 10$K, the spacing $\delta E\approx
100$mK, this is typical for semiconductor QD. It is remarkable
that that the Andreev levels do not depend on microscopic and
geometric details of the constriction by means other than
conductance and transmission amplitudes. This fact considerably
simplifies the fabrication of Andreev dots with discrete Andreev
levels.

In the ground state of the dot, quasiparticles occupy Andreev
levels with negative energy. The $\varphi$-dependent part of the
ground state energy reads
$E_0=1/2\sum_{n\sigma}\varepsilon_{n\sigma}
\Theta(-\varepsilon_{n\sigma})$ where we sum over channel and spin
index; $\Theta(x)=1$ for $x>0$ and zero for $x\leq0$. The factor
$1/2$ comes from the double-counting mentioned.\cite{17_1/2} If
spin splitting of Andreev levels is smaller than $\delta E$, the
ground state has zero spin, since both components of the spin
dublet are occupied by quasiparticles. In an excited state of the
AQD, some Andreev levels with positive energy are populated. Let
us concentrate on a given transport channel where there are two
such Andreev levels corresponding to two spin directions. One
quasiparticle fills either level, the AQD has spin-1/2. Second
 quasiparticle fills the level with the opposite spin
 resulting in an excited spin-singlet state. (Fig. 1a)
The spin-1/2 state of an AQD with the lowest energy
(that corresponds
to the most transparent transport channel)
is of particular interest
because it is very stable.
The transition to ground state require the 1/2 change of
spin. This means that a quasiparticle must either leave or
enter the AQD. The probabilities of these processes
contain exponentially small factors $\exp({-\Delta/k_BT})$,
this means that at zero temperature the AQD would remain
in spin-1/2 forever.
 The physics involved is very
similar to well-known parity effect in superconducting grains
\cite{18Averin-Nazarov}. Thus it is possible to preserve the system
for a long time in  this spin-1/2 state.

\begin{figure}[htb]
\includegraphics[width=80mm]{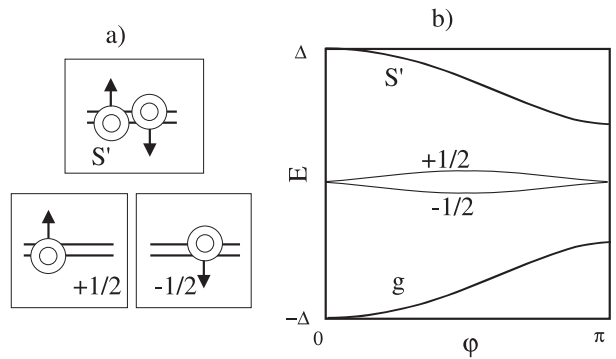}
\caption{\label{fig1}  a) Excited states of Andreev
quantum dot in a given transport channel.
 b) The energies of AQD states corresponding to a given transport
level versus phase difference. The lower and upper
curve correspond to ground state and excited singlet respectively.
The middle curves correspond to two spin-1/2 excited states, spin
splitting being due to spin-orbit interaction.}
\end{figure}

How to \textit{\textbf{set}} the AQD to spin-1/2 state?
Possibilities include microwave irradiation \cite{19Lundin} and
quasiparticle injection \cite{20Wendin}. We concentrate on the
first possibility. Let us assume that the irradiation frequency
$\omega$ satisfies $\Delta+\varepsilon_0<\hbar\omega<2\Delta$.
Under these conditions, the absorption of the irradiation quanta
takes place in the constriction only. If initial state is
the ground state, each absorption generates two quasiparticles.
There are two possibilities: 1) both
quasiparticles appear in bound Andreev states; 2) one
quasiparticle appears in a bound state whereas another one
acquires energy~$>\Delta$ and gets to the extended state
(photoemission); the latter quasiparticle leaves the AQD almost
immediately and never comes back.
Next absorption process can occur in excited state, in this
case the energy balance allows extra processes: 1) two
quasiparticles leave the dot, 2) one quasiparticle in a bound
state is excited over the energy barrier and leaves the dot.
In any case, the processes of the
first type are irrelevant not changing the parity of the AQD.
Processes of the second type switch the AQD between
states with even and odd number of quasiparticles.
If the irradiation lasts
long enough for many processes of the second type to occur,
the AQD is in the state with half-integer spin with 50 \% probability.
Let us now switch off the irradiation.
If there is an even number of quasiparticles in the AQD, the
subsequent energy relaxation will drive the system to the ground
state. For an odd number of particles, the relaxation will result
in a \textit{stable} single quasiparticle occupying the lowermost
Andreev level, the lowermost spin-1/2 state we are after.

How to \textit{\textbf{detect}} the spin-1/2 state? It is
important that the superconducting current in the constriction is
different for the states involved, since their $\varphi$-dependent
energies are different. Here we concentare
on a channel $n$ and count energy from the ground state energy
of quasiparticles in all other channels, $E_g+|\varepsilon_n|$.
The energies then are $-|\varepsilon_n|$, $0$, and
$|\varepsilon_n|$ for ground, spin-1/2, and excited singlet state
respectively.(Fig. 1b)
and the superconducting
current equals to $I=e\partial_\varphi E/\hbar$. The
change from the ground to spin-1/2 state is therefore manifested
as a change of supeconducting current by a value of $\delta
I\equiv e\partial_\varphi \varepsilon_{n\sigma}/\hbar$.
The detection of such current jumps in
superconducting constrictions would amount to the direct
experimental observation of the spin-1/2 state.

How to detect \textit{\textbf{spin}} in the spin-1/2 state? An
important advantage of an AQD is that its spin state affects
the superconducting current, the latter being detected
and/or measured. This is due to spin-orbit splitting of Andreev
levels. Generally, one expects such splitting given the fact that
the supeconducting phase difference changes sign under time
reversal so that Kramers theorem  does not hold. A
confusing circumstance is that the Kramers theorem does hold in
the universal limit of short constriction considered. To get
spin-orbit splitting, one considers extra perturbative corrections
of the first order in $\tau_{_{\rm flight}}\Delta/\hbar$. The
calculation based on the scattering matrix approach
\cite{16Beenakker} with the use of Eq.\eqref{BdG} yields
the following effective Hamiltonian for the spin-spitting
in question:
\begin{gather}
\label{E_SO}
E_n^{(SO)}=\Delta
(\boldsymbol{\alpha}_n\cdot\boldsymbol{\sigma})\sin(\varphi)(\tau_{_{\rm
flight}}\Delta/\hbar),
\end{gather}
$\boldsymbol{\sigma}$ being the pseudovector operator of spin.
 Here a dimensionless pseudovector
$\boldsymbol{\alpha}_n$ is a property of a given Andreev level not
depending on $\varphi$; it is proportional to the spin-orbit
constant that routinely contains a nuclear charge of the material
$Z$, $|\boldsymbol{\alpha}|\backsimeq Z (e^2/\hbar c)$. For
example, in a quasi-ballistic SNS junction $\boldsymbol{\alpha}$ is
directed along the vector product of the quasiparticle momentum
$\bf p$ ($|\mathbf{ p}|\approx p_F$) and the direction of the
electric (crystal) field $\bf E$ for $\varphi>0$:
$\boldsymbol{\alpha}\upuparrows \mathbf{p}\times \mathbf{E}$; but
when $\varphi<0$, $\bf p$ changes its direction to the opposite
and $\boldsymbol{\alpha}\upuparrows -\mathbf{p}\times \mathbf{E}$.
This change of the direction of the quasiparticle momentum
(Andreev state ``chirality'') with the sign of $\varphi$ is the
reason why the Andreev level spin-splitting in Eq.\eqref{E_SO} has
rather unusual odd dependence on $\varphi$ [this is general
property of spin-orbit Andreeev level splitting]. The technical model
to derive
 Eq.\eqref{E_SO} was a one-channel conductor
 with  two scatterers separated by a distance corresponding
 to $\tau_{\rm flight}$. Both scattering matrices contained
 spin-orbit part. The model showed
that in not very short constrictions ($\tau_{_{\rm
flight}}\Delta/\hbar\simeq 1$) spin-orbit level splitting aquired
more complicated dependence on $\varphi$ and could be of the
order of $\Delta$ provided spin-orbit scattering was
comparable with orbital one.
Thus the splitting can be huge:
for a material with heavy nuclei $\boldsymbol{\alpha}$ can
become of the order of unity. To show how big can be the effect,
let us consider an AQD embedded into a superconducting loop with
self-inductance $L$. The spin produces an extra superconducting
current $I_\sigma=e\sigma\partial_\varphi
E^{(SO)}\hbar$ and an extra magnetic
moment per spin thereby. Let us estimate the maximum possible
value of this magnetic moment. For this, we set  $\tau_{_{\rm
flight}}\simeq \Delta/\hbar$ ,$|\boldsymbol{\alpha}| \simeq 1$, and
concentrate on a resistive constriction $R \backsimeq R_Q$ [so that
the critical current is of the order of $e\Delta/\hbar$]. The
inductance $L$ of a typical SQUID loop does not exceed $1cm$
\cite{Likharev} thus the extra magnetic moment $\delta M\lesssim
(e\Delta/\hbar) L^2/c\sim 10^{12}\mu_B$.

Let us give a simple example of spin
\textit{\textbf{manipulation}} in the AQD. For this, we need to
invoke  Zeeman splitting of  Andreev level in magnetic field
$\sigma E^{(Z)}$ in addition to spin-orbit
splitting $\sigma E^{(SO)}_{n}$. This allows us to
control the direction of spin quantization by the magnetic field,
this possibility being absent if we work with spin-orbit only. To
achieve comparable $E^{(Z)}$ and
$E^{(SO)}$, the magnetic field should be almost
"in plane"  not affecting the flux in the SQUID loop
and thus $\varphi$. Let us assume that we manage to
achieve this so that the quantization axis of spin may deviate
substantially from $\boldsymbol{\alpha}$. The spin wave function
is therefore a coherent mixture of the states $|\uparrow\rangle$,
$|\downarrow\rangle$ with spin parallel or antiparallel to
$\boldsymbol{\alpha}$ respectively. Quite generally,
$\Psi=a|\uparrow\rangle+b|\downarrow\rangle$, where
$|a|^2+|b|^2=1$. If there is no in-plane field, the system is in
the ground state $|\downarrow\rangle$,spin quantization axis being
parallel to $\boldsymbol{\alpha}$. Now let us switch on the
in-plane field. The Hamiltonian governing dynamics of the wave
function will thus become : $H=g\mu_B
(\boldsymbol{\sigma}\cdot\mathbf{ H})+\hat E_{(SO)}$.
Let us assume the simplest form of the resulting
Hamiltonian: $H=E^{(Z)} \hat \sigma_y +E^{(SO)}\hat\sigma_z$,
$E^{(Z)}\sim E^{(SO)}$. The
wave function will then evolve according to
\begin{eqnarray*}
\Psi(t) =|\downarrow\rangle\left(\cos (\Omega t)+i\sin(\Omega t)
({E^{(SO)}}/{\hbar\Omega})\right)-
\\
-|\uparrow\rangle \sin(\Omega t)({E^{(Z)}}/{\hbar\Omega}),
\end{eqnarray*}
where $\hbar\Omega=\sqrt{(E^{(SO)})^2+(E^{(Z)})^2}$  is
the frequency of Rabi oscillations. It is
important to note that these oscillations can be readily
\textit{\textbf{detected}} since they produce an alternating
current
\begin{gather}
I_a(t)=2\sin^2(\Omega
t)\left(\frac{E^{(Z)}}{\hbar\Omega}\right)^2e\partial_\varphi
E^{(SO)}/\hbar\,.
\end{gather}
Another way of manipulation is readily borrowed from the quantum
optics: if the in-plane magnetic field oscillates with the
resonant frequency $E_{SO}$, a significant manipulaton
effect can be achieved
even if  $E^{(Z)} \ll E^{(SO)}$ \cite{book:Mandel_Wolf}. A
general unitary transformation of the spin-wave function can be
performed exposing the junction to the time-dependent $E^{(Z)}$,
quite similar to many other solid-state implementations of the
qubits.

Let us discuss now how the Andreev quantum dots can be utilized
for universal \textit{\textbf{quantum computations}}. An AQD in
the spin-1/2 state would be a qubit. We have discussed above how to
manipulate the spin of a single AQD. This is how the single-qubit
operations can be performed.  A quantum computation algorithm
should involve two-qubit operations as well. An important
theoretical result \cite{3Ekert-Jozsa} establishes that the XOR
operation with two qubits along with single-qubit operations forms
a minimum set that is sufficient to build up an arbitrary complex
universal quantum computer. The XOR operation does the following:
given two qubits in the states $|x\rangle$, $|y\rangle$, it leaves
the $|y\rangle$ state unchanged if $|x\rangle=|\uparrow\rangle$,
while flipping it when $|x\rangle=|\downarrow\rangle$. So we
concentrate on a possible realization of XOR operation for two
AQD's.

\begin{figure}[htb]
\includegraphics[width=80mm]{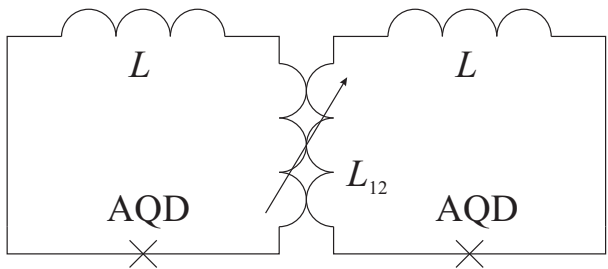}
\caption{\label{fig2} The interaction between AQD qubits is
organized by means of variable inductive coupling. This allows for
further arraying. The circuit shown can be used as an XOR gate.}
\end{figure}

The basic idea is to organize the interaction between AQD's via
inductive coupling between SQUID loops containing these AQD's [see
Fig.~\ref{fig2}].
The two-qubit operations are performed by varying the mutual
inductance between two given loops. Indeed, in this case the
interaction between two AQD's can be described by a simple
hamiltonian  $H=L_{12}I_1 I_2$, $L_{12}$ being the mutual
inductance, $I_{1(2)}$ standing for operators of  currents in
corresponding loops. Owing to spin-orbit interaction, each current
depends on spin state of corresponding AQD:
$I_{1,2}=I_{1,2}^{(0)}+I_{1,2}^{s}s_{1,2}^z$. Here we choose
$z$-axis in the space of each spin in the direction of
corresponding pseudovector $\boldsymbol{\alpha}$. This is to
stress the following circumstance: although the pseudovectors
$\boldsymbol{\alpha}$ in different AQD's may differ, there is
always only one component of spin that is reflected in the current
and therefore takes part in the interaction. So that, the relevant
part of the Hamiltonian can be written as
\begin{gather}
\label{three} H=H_1s_1^z+H_2s_2^z+H_{12}s_{1}^zs_2^z.
\end{gather}

This simple Ising-type form of Hamiltonian brings us to the
old-fashioned but solid "optical" quantum computer
\cite{21Lloyd}. In this approach, the one-bit operations are
performed at $H_{12}=0$ by pulses at resonant frequencies
$H_1/\hbar$ or $H_2/\hbar$, the pulse duration being tuned to
achieve the spin flip. The XOR operation is performed at
$H_{12}\neq 0$ by the same pulse with frequency $(H_1+H_2)/\hbar$.
An alternative way is to use non-oscillating pulses of
$H_{12}/\hbar$. Such pulses would shift phases of two states with
antiparallel spins with respect to the phases of the states with
parallel spins thus realizing ``quantum phase shift gate''
\cite{22Barenco}. The XOR operation can be performed by combining
two such phase shifts with two rotations of the target spin.

This approach of organizing two-qubit interactions has two
important practical advantages. First, in contrast to other
spin-based proposals, the interaction does not have to be
organized at microscopic level. To exaggerate, one can use
inch-scale transformers to vary inductive coupling between the
AQD's. To make a practical suggestion, one can use the
well-developed techniques of SQUID circuitry
\cite{23Turchette,24Mukhanov} to couple, array, bias, and measure
many AQD qubits. Second advantage is the simple Ising form of the
resulting interaction that prevents undesired phase shifts and
simplifies design of complicated quantum circuits.

To conclude, we analyze prospectives of Andreev quantum dots for
spin manipulation and quantum computing. Our theoretical results
seem to be promising enough to launch detailed experimental
investigations and design efforts in this direction.

We wish to thank RFBR (projects No. 03-02-06259, 03-02-16677, and
03-02-17494), the Netherlands Organization for Scientific Research
(NWO), the Swiss NSF, the programs of the Russian Ministry of
Science: Mesoscopic systems and Quantum computations and the
program of the leading scientific schools support.


\begin{thebibliography}{99}
\bibitem{1Vion-Devoret} D. Vion {\it
et al}., Science \textbf{296}, 886 (2002)

\bibitem{2Yu-Jozsa} Y. Yu {\it
et al}., Science \textbf{296}, 889 (2002)

\bibitem{3Ekert-Jozsa}
A. Ekert, R. Jozsa, Rev. Mod. Phys. \textbf{68}, 733 (1996).

\bibitem{4Loss-DiVincenzo}
D.Loss, D.P. DiVincenzo, Phys. Rev. A \textbf{57}, 120 (1998).

\bibitem{5Recher-Levy}
P. Recher, D. Loss, J. Levy, Cond-mat/0009270.

\bibitem{6Shnirman-Hermon}
A. Shnirman {\it et al}., Phys. Rev. Lett. \textbf{79}, 2371
(1997)

\bibitem{7Averin}
D.V. Averin, Solid State Commun. \textbf{105}, 659 (1998).

\bibitem{8Mooij-Lloyd}
J.E. Mooij {\it et al}., Science \textbf{285}, 1036 (1999).

\bibitem{9Kane}
B.E. Kane, Nature \textbf{393}, 133 (1998).

\bibitem{10Makhlin-Schnirman}
Yu. Makhlin {\it et al}., Rev. Mod. Phys. \textbf{73}, 357 (2001).

\bibitem{11Ioffe-Blatter}
L.B. Ioffe {\it et al}., Nature \textbf{398}, 679 (1999).

\bibitem{12Nakamura-Tsai}
Y. Nakamura {\it et al}., Nature \textbf{398}, 786 (1999).

\bibitem{13Ioffe}
L.B. Ioffe {\it et al}., Nature \textbf{415}, 503 (2002).

\bibitem{14Recher}
P. Recher {\it et al}., Phys. Rev. Lett. \textbf{85}, 1962 (2000).

\bibitem{Andreev} A.F.\ Andreev, \newblock Sov.\ Phys.\ JETP {\bf 19}, 1228
(1964) 

\bibitem{dG}  P.G. de Gennes, ``Superconductivity of Metals and
Alloys'', chapters 5, 8, W.A. Benjamin, inc., New York - Amsterdam
1966.

\bibitem{Landau3} L.D. Landau and E.M. Lifshitz, in
\textit{Quantum Mechanics}, Course in Theoretical Physics Vol. 3
(Pergamon Press, Oxford, 1977).

\bibitem{17_1/2}
Bardeen {\it et al}., Phys. Rev. 187, 556 (1969).
Communications with our colleagues have convinced us
that this basic point has to be discussed in this article.


\bibitem{16Beenakker}
C.W.J. Beenakker, Phys. Rev. Lett. \textbf{67}, 3836 (1991).


\bibitem{15Imry}
Y. Imry, Introduction to mesoscopic physics, Oxford University
Press, 1997.

\bibitem{18Averin-Nazarov} D.V. Averin {\it
et al}., Phys. Rev. Lett. \textbf{69}, 1993 (1992); M.T. Tuominen
{\it et al}., Phys. Rev. Lett. \textbf{69}, 1997 (1992).


\bibitem{19Lundin}
N.I. Lundin {\it et al}., Superlattices and Microstructures
\textbf{25}, 937 (1999)

\bibitem{20Wendin}
G. Wendin {\it et al}., Superlattices and Microstructures
\textbf{25}, 983 (1999).



\bibitem{Likharev} K.\,K.\,Likharev, {\it
Dynamics of Josephson Junctions and Circuits} (Gordon and Breach
Science Publishers, Amsterdam, 1991).



\bibitem{book:Mandel_Wolf} L.\ Mandel and E.\ Wolf,
   {\it Optical Coherence and Quantum Optics}, (Cambridge
   Univ. Press, Cambridge, UK, 1995).



\bibitem{21Lloyd}
S.A. Lloyd, Science \textbf{261}, 1569 (1993).

\bibitem{22Barenco}
A. Barenco {\it et al}., Phys. Rev. Lett. \textbf{74}, 4083 (1995)

\bibitem{23Turchette}
Q.A. Turchette {\it et al}., Phys. Rev. Lett. \textbf{75}, 4710
(1995)

\bibitem{24Mukhanov}
O.A. Mukhanov {\it et al}., Physica C \textbf{368}, 196 (2002)


\end{thebibliography}
\end{document}